\documentclass[useAMS,usenatbib]{mn2e}
\usepackage{graphicx}
  
\def\refjnl#1{{\rm#1}}
\def\apj{\refjnl{ApJ}}                 
\def\mnras{\refjnl{MNRAS}}             

\def\be{\begin{equation}} 
\def\ee{\end{equation}}

\def\gsim{\lower.5ex\hbox{\gtsima}} 
\def\lsim{\lower.5ex\hbox{\ltsima}} 
\def\gtsima{$\; \buildrel > \over \sim \;$} 
\def\ltsima{$\; \buildrel < \over \sim \;$} 
\def\prosima{$\; \buildrel \propto \over \sim \;$} 
\def\gsim{\lower.5ex\hbox{\gtsima}} 
\def\lsim{\lower.5ex\hbox{\ltsima}} 
\def\simgt{\lower.5ex\hbox{\gtsima}} 
\def\simlt{\lower.5ex\hbox{\ltsima}} 
\def\simpr{\lower.5ex\hbox{\prosima}} \def\la{\lsim} \def\ga{\gsim} 
  
 \def\gtsima{$\; \buildrel > \over \sim \;$} 
\def\ltsima{$\; \buildrel < \over \sim \;$} 
\def\gsim{\lower.5ex\hbox{\gtsima}} 
\def\lsim{\lower.5ex\hbox{\ltsima}} 
\def\simgt{\lower.5ex\hbox{\gtsima}} 
\def\simlt{\lower.5ex\hbox{\ltsima}} 
\def\simpr{\lower.5ex\hbox{\prosima}} 
\def\la{\lsim} 
\def\ga{\gsim}

\def\lya{Ly$\alpha$~} 
\def\fesc{$f_{esc}$~}
\def\fescrel{$f_{esc,rel}$~}

\def\E3{{\cal E}_{\rm g}^{III}}

\title[Reionization in the pre-ELT era]{Probing ionizing radiation of $L\lsim 0.1L^{*}$
star-forming galaxies at $z\gsim 3$ with strong lensing} 
\author[Vanzella et al.]{Eros Vanzella$^{1}$\thanks{E-mail:vanzella@oats.inaf.it}, Mario Nonino$^{1}$, 
Stefano Cristiani$^{1}$, Piero Rosati$^{2}$, Adi Zitrin$^{3}$, \and  Matthias Bartelmann$^{3}$,
Andrea Grazian$^{4}$, Tom Broadhurst$^{5,6}$, Massimo Meneghetti$^{7}$, \and Claudio Grillo$^{8}$\\\\\\
$^{1}$ INAF Osservatorio Astronomico di Trieste, Via G.B.Tiepolo 11,  34131 Trieste, Italy\\
$^{2}$ European Southern Observatory, Garching bei M¨unchen, Germany\\
$^{3}$ Institut f\"ur Theoretische Astrophysik, Universit\"at Heidelberg, Heidelberg, Germany\\
$^{4}$ INAF Osservatorio Astronomico di Roma, Via Frascati 33,00040 Monteporzio (RM), Italy\\
$^{5}$ Theoretical physics, University of the Basque Country, Bilbao 48080,  Spain\\
$^{6}$ Ikerbasque, Basque Foundation for Science, Alameda Urquijo, 36-5 Plaza Bizkaia 48011, Bilbao, Spain \\
$^{7}$ INAF - Osservatorio Astronomico di Bologna, via Ranzani 1, 40127, Bologna, Italy\\
$^{8}$ Dark Cosmology Centre, Niels Bohr Institute, University of Copenhagen, Juliane Mariesvej 30, DK-2100 Copenhagen, Denmark}
\begin{document} 
 
\date{} 
 
 
\maketitle 
 
\label{firstpage} 
\begin{abstract} 
We show the effectiveness of strong lensing in the characterisation of
Lyman continuum emission from faint $L\lsim 0.1L^{*}$ star-forming
galaxies at redshift $\gsim 3$.  Past observations of $L\gsim L^{*}$
galaxies at redshift $\gsim 3$ have provided upper limits of the
average escape fraction of ionising radiation of \fesc $\sim$ 5\%.
Galaxies with relatively high \fesc ($>10\%$) seem to be particularly
rare at these luminosities, there is therefore the need to explore fainter
limits.  Before the advent of giant ground based telescopes, one
viable way to probe \fesc down to $0.05-0.15L^{*}$ is to exploit
strong lensing magnification.  This is investigated with Monte Carlo
simulations that take into account the current observational
capabilities. Adopting a lensing cross-section of $10~arcmin^{2}$
within which the magnification is higher than 1 (achievable with about
4-5 galaxy clusters), with a U-band survey depth of $30$($30.5$) (AB,
1$\sigma$), it is possible to constrain \fesc for $z\simeq3$
star-forming galaxies down to 15(10)\% at 3-sigma for $L<0.15L^{*}$
luminosities.  This is particularly interesting if \fesc increases at
fainter luminosities, as predicted from various HI reionization
scenarios and radiation transfer modelling.  Ongoing observational
programs on galaxy clusters are discussed and offer positive prospects
for the future, even though from space the HST/WFC3 instrument
represents the only option we have to investigate details of the
spatial distribution of the Lyman continuum emission arising from $z
\sim 2-4$ galaxies.
\end{abstract}


\section{Introduction}

Recent works suggest that at $z>3$
star-forming galaxies are the leading candidates for the production of ionising photons
(e.g., Kuhlen \& Faucher-Giguere 2012;
Haardt \& Madau 2011; Ciardi et al. 2012).
However, the mechanisms regulating
the escape fraction of ionising radiation (\fesc) from galaxies are still unknown.
In particular it is not clear if \fesc evolves with cosmic time, and at fixed redshift, if
it is luminosity dependent.
The complexity in modelling of galaxy evolution and the inclusion 
of radiative transfer prescriptions make the predictions on \fesc
very uncertain, and opposite results are often obtained from simulations. 
Moreover, the predictions on \fesc show a large variation from galaxy to galaxy,
between 0.01 to nearly 1, as a result of differences in the hydrogen distribution
(e.g., Gnedin et al. 2008; Yajima et al. (2011); Fernandez \& Shull (2011), and references therein).
From an observational point of view, the current situation is also far from clear.
Various observations have provided significant upper limits on Lyman continuum
(LyC) emission in the redshift range 1-4, i.e., \fesc smaller than 5-10\%
(Vanzella et al. 2012, V12, and references therein).
In particular Vanzella et al. 2010, V10b, and Boutsia et al. (2011) 
reported upper limits on \fesc of $5\%$ for $L\gsim L^{*}$ LBGs at redshift 3--4.
Very few galaxies with possible LyC detection have been reported in the literature
(Shapley et al. 2006; Iwata \& Inoue 2009; Nestor et al. 2011).
The fact that bright sources with high \fesc ($> 10\%$) are rare
could support the interpretation that \fesc increases at lower luminosities.
Indeed, recent radiative transfer calculations coupled with cosmological
simulations show that low luminosity and dwarf galaxies
have high \fesc ($>20-40\%$) and are the major contributors to the 
ionizing background at redshift $3<z<6$ (e.g., Wise \& Cen 2009;
Razoumov \& Sommer-Larsen 2010; Yajima et al. 2011;
Fernandez \& Shull 2011; but see Gnedin et al. (2008) for different results).
The investigation of faint (and/or low mass) star-forming galaxies
at moderate redshift ($z\sim 3-4$) is therefore crucial to probe
ionisation regimes and conditions that would be in place during the 
reionization epoch ($z>7$). It is worth noting that 
the reionization at $z = 7-10$
requires an average \fesc much higher than what is observed
for relatively bright and lower redshift galaxies ($z<4$)
(Robertson et al. 2010; Bouwens et al. 2010; Fernandez \& Shull 2011).

While the direct measurements of escaping LyC 
photons are prohibitive during the epoch of reionization, they are
still accessible at $z\lsim4$.
The investigation of sub-$L^{*}$ galaxies requires extremely deep U-band 
surveys ($U>30$). However, even if they were available,  spectroscopic redshift 
measurements for $L<<L^{*}$ galaxies (e.g., r-band$>27$) are very
challenging with current facilities.
In this respect, before the advent of the ELT and JWST telescopes, 
strong lensing magnification offers a viable way to explore faint limits
in luminosity (and mass).

\section{The \fesc and strong lensing}
\label{basics}

The {\it relative} fraction of escaping LyC photons (at $\lambda < 912$\,\AA)
relative to the fraction of escaping non-ionising ultraviolet (1500\,\AA)
photons is defined as (Steidel et al. 2001): 
\begin{equation} 
f_{\rm esc,rel} \equiv \frac{(L_{1500}/L_{LyC})_{\rm int}}{(\frac{\mu ~ F_{1500}}{\mu ~ F_{LyC}})_{\rm
obs}}  \exp(\tau^{\rm IGM}_{LyC}), \label{f_esc_rel} \label{eq:fesc_rel}
\end{equation}

\noindent where $(F_{1500}/F_{LyC})_{\rm obs}$, $(L_{1500}/L_{LyC})_{\rm int}$ and $\tau^{\rm IGM}_{LyC}$
represent the observed 1500\,\AA/LyC flux density ratio,
the intrinsic 1500\,\AA/LyC luminosity density ratio, and
the line-of-sight opacity of the IGM for LyC photons, respectively.
It is worth noting that the two quantities $(F1500)_{\rm  obs}$
and $(F_{LyC})_{\rm obs}$ are measured in the same spatial
(i.e., physical) region, where the ionising and non-ionising radiation arise (V12).
If the dust attenuation $A_{1500}$ is known, 
$f_\mathrm{esc,rel}$ can be converted to the absolute $f_\mathrm{esc}$ as 
$f_\mathrm{esc} = 10^{-0.4A_{1500}} f_\mathrm{esc,rel}$
(e.g., Siana et al. 2007).
The term $\mu$ is the magnification factor provided by a lens (e.g., a cluster
of galaxies) and applies to both flux densities observed at two wavelengths
(we leave the parameter explicitly in the expression).
We assume for simplicity it does not vary substantially 
within the isophote of the sources.
Therefore, given the achromatic nature of lensing, \fesc is independent from $\mu$.
The two quantities \fesc and \fescrel are equal if $A_{1500}=0$. Conservatively, 
in the following we perform calculations assuming $A_{1500}=0$, 
the limits probed on \fesc are deeper if $A_{1500}>0$, e.g., they are 
halved if $A_{1500}=0.6$ (V10b).
If not specified, we assume an intrinsic luminosity density ratio  
$(L_{1500}/L_{LyC})_{\rm int} = 7$ (Siana et al. 2007; V10b).
In Sect.~3.2, the IGM transmission of Inoue et al. (2008,2011) is convolved
with the adopted U-band filter.
We perform calculations for three U-band filters:
F336W ({\it F336W}), u-band Str\"omgren ({\it U-strom}) and U-special 
available at the Large Binocular Telescope ({\it LBC-U}) 
(in general the results do not change if other filters with
similar widths and wavelength coverages are considered).
It is clear from Eq.~(\ref{eq:fesc_rel}) that high magnitude contrasts 
between the LyC and the 1500\AA~rest-frame provide strong constraints 
on \fesc. 
We can rearrange Eq.~\ref{eq:fesc_rel} to give an estimate of the expected
magnified flux at wavelengths smaller than the Lyman limit
as a function of \fesc and magnification factor $\mu$ 
(i.e., $F_{\rm LyC}^{Lens}=\mu F_{\rm LyC}$):

\begin{equation}
F_{\rm LyC}^{Lens} =  \left ( \frac{L\lambda_{\rm rest}}{L_{1500}} \right )_{\rm int}  \frac{f_{\rm esc} \times (F_{1500} \times \mu)_{\rm obs}}{e^{\tau^{\rm IGM}_{\lambda}}} \times 10^{0.4\times A1500}. \label{f900_obs}
\label{eq:f900_obs}
\end{equation}

\noindent As discussed in Inoue \& Iwata (2008) and V10b,
the stochastic nature of the IGM absorption introduces large uncertainties in the estimate 
of \fesc for a single line of sight.
Stacking many galaxies among different lines of sight
provides strong limits and reduces the variance due to IGM attenuation.
Indeed, a deep U-band survey can provide firm constraints 
for $L\gsim L^{*}$ galaxies down to few percent of \fesc by stacking 
tens of objects. This has been performed by V10b with deep
VLT/VIMOS U-band imaging ($\approx 30AB$ at $1\sigma$) of $z\gsim3.4$ LBGs
(see also Boutsia et al. (2011)).

For $L<<L^{*}$ galaxies the situation becomes challenging.
One reason is that the knowledge of the redshift 
is necessary to fix the observed wavelength position of the Lyman limit.
Even though a deep U-band survey (e.g, mag-U$\simeq 30$ at 1-sigma) 
can still probe \fesc down to 20\% by stacking about thirty 0.1$L^{*}$ galaxies, 
the redshift confirmation is not practicable with current facilities,
or it is feasible only for a sub-class of sources
like the \lya--~emitters.
The lensing magnification greatly facilitates
the redshift measurement also for faint galaxies.
Additionally it allows 
the investigation of \fesc down to stronger limits, by
the U-band stacking and/or individual LyC detections.
In the following we perform Monte-Carlo (MC) simulations to
derive predictions for \fesc as a function of the U-band depth, 
lensing magnification and lensing cross-section.
In what follows we adopt $r$ and $R$ to indicate the r-band magnitude 
($\sim$ 1500\AA~rest-frame) for non-magnified and magnified sources, respectively.

\begin{figure}
\begin{center}
\includegraphics[angle=0,width=3.25in]{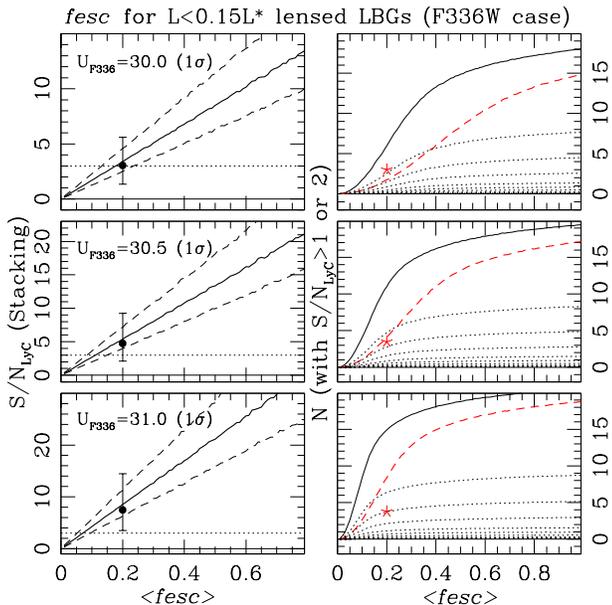}
\end{center}
\scriptsize
\caption{
\label{MC}
MC simulations for magnified LBGs at $3<z<3.5$
with intrinsic $L< 0.14L^{*}$ ($r>26.5$) and adopting a lensing cross-section of $10~arcmin^{2}$
(see text for details). Top, middle and bottom panels show the results of our MC simulations
for U-band depths of 29, 30 and 31 (1-sigma), respectively.
{\bf Left panels:} the expected S/N ratio in the LyC of the stacked fluxes
(median and central 68\% interval, solid and dashed lines, respectively).
Filled circles are the S/N ratio (median and central 68\% interval)
in the ``On/Off case'' with 20\% of the galaxies with \fesc=1 and 80\% with \fesc=0.
Horizontal dotted lines mark S/N=3.
{\bf Right panels:} number of direct detections in the LyC at $S/N>1$ 
(black solid line) and $S/N>2$ (red dashed line).
The dotted black lines report the number of galaxies in the faint magnitude bins from which
they migrate to brighter apparent magnitudes ($R<26$).
The black solid line is the sum of the dotted lines.
From top to bottom the dotted lines refer to the magnitude bins $26.5<r<27.0$, $27.0<r<27.5$, etc.
The red stars mark the expected number of direct detections at $S/N>2$ in the
``On/Off case''.}
\end{figure}

\section{Magnified galaxies: MC simulations}
We want now to estimate the feasibility of probing \fesc down to $<20\%$ for
$L\lsim 0.1L^{*}$ LBGs. To this aim, 
from Eq.~\ref{eq:f900_obs} it is possible to calculate 
as a function of \fesc and the U-band depth the number of sources detectable in 
the LyC above a certain threshold and the signal to noise ratio in their LyC stack. 
We therefore perform MC simulations similar to those 
described in V10b by assuming reasonable distributions 
for the quantities involved in Eq.~\ref{eq:f900_obs}.

\subsection{The expected number of magnified LBGs}

The probability that a source galaxy is magnified by more than $\mu$ 
can be expressed in terms of the probability density $P(\mu)$: 
$P(>\mu) = \int_{\mu}^{\infty} P(\mu)d\mu$ where $P(\mu)=-dP(>\mu)/d\mu$.
An interesting property of the lensing probability in the source plane 
is that $P(>\mu) \propto \mu^{-2}$ and therefore $P(\mu) \propto \mu^{-3}$ 
for $\mu>>1$ as can be shown in particular cases and argued 
to be true in general (see Fig.~9 of Lima et al. 2010; Schneider et al. 1992).
A maximum magnification ($\mu_{max}$) is imposed by the size of the source galaxies
that we assume to have an intrinsic half light radius not smaller 
than 0.2kpc (assuming circular shape), i.e., $\mu_{max}= 400$.

Each galaxy cluster produces a region within which the magnification in the source plane
is larger than $\mu_{min}$. This area is an effective cross-section for lensing statistics.
For our purposes, it is enough to perform a simple calculation for the cross 
section $\sigma_{lens}$:

\begin{equation}
\sigma_{lens} \simeq \frac{1}{\mu_{min}^{2}} \sum_{i=1}^{N_{clust}} \pi \theta_{E,i}^{2} ~~~~ (arcmin^{2})
\label{sigma}
\end{equation}

\noindent where $\theta_{E,i}$ is the Einstein radius of a given cluster.
Lenses with the largest $\theta_{E}$ are the best ``cosmic telescopes'', 
in particular we consider in our simulations the case with $\sigma_{lens}=10~arcmin^{2}$
and $\mu_{min}=1$, 
corresponding approximately to 4-5 massive 
galaxy clusters (see Sect. 4), the results can be linearly rescaled to other areas.

Adopting the $z\sim3$ luminosity function parameters of Reddy \& Steidel (2009)
($\alpha=-1.73$, $M^{*}_{AB}=-20.97$, $\phi^{*}=1.71 \times 10^{-3} Mpc^{-3}$), we
can calculate the expected number of star-forming galaxies within the chosen area
and redshift interval {\it dz}. 
A further important constraint is that the magnification must produce an apparent magnitude $R$ brighter 
than the spectroscopic limit for redshift measurements ($R<m_{spec}$), 
necessary to fix exactly the wavelength position of the Lyman limit. We adopt $m_{spec}=26$ that 
corresponds to a relatively deep spectroscopic survey (e.g., Vanzella et al. 2009).
Together with the limit on $\mu$ ($< 400$) defined above, the chosen  $m_{spec}$ implies 
a limit on the integration of the luminosity function, that is $r=32.5$ (i.e.,
sources with $r>32.5$ would require $\mu>400$ to have $R<26$).
It turns out that within an area $\sigma_{lens}=10~arcmin^{2}$ and redshift interval $3.0-3.5$,
$\sim2300$ LBGs are expected in the magnitude 
range $26.5<r<32.5$, 30 of which have $R<m_{spec}=26$ (we refer to this sample as $N_{spec}=30$).
Clearly, the intrinsically faint galaxies are the rarer cases
accessible from ground-based spectroscopy, being sources
that need large magnification $\mu$.
We anticipate here that the constraints we discuss below on \fesc
are mainly dictated by sources belonging to the magnitude bin $26.5<r<28$
(i.e., $L=0.05-0.16L^{*}$, on average 24 of the 30 galaxies).

\subsection{MC simulations of $L\lsim0.1L^{*}$ LBGs}
\begin{figure}
\begin{center}
\includegraphics[angle=0,width=3.25in]{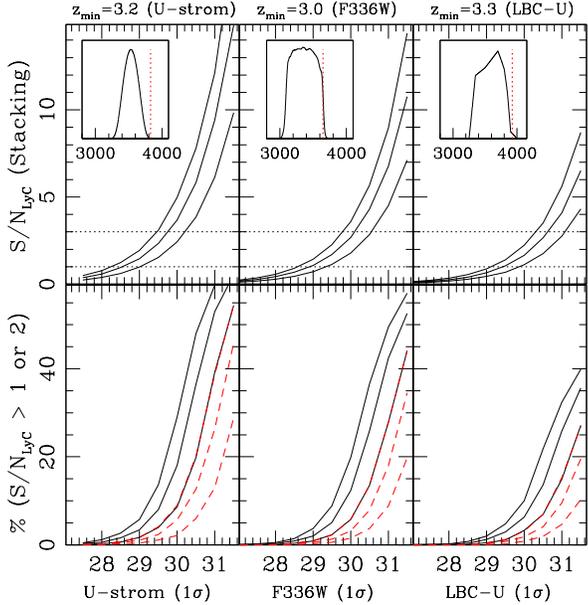}
\end{center}
\scriptsize
\caption{
\label{MC2}
Signal to noise in the LyC of the stacking (top panels) and the fraction (\%) of LyC 
detections with S/N higher than 1 and 2 (bottom panels) are shown as a function of the
U-band depth. Three different U-band filters have been used. 
In the three cases $\sigma_{lens}=10~arcmin^{2}$ and $N_{spec}\simeq30$. 
In all panels the three curves have been calculated
 for \fesc=0.2, 0.15 and 0.10 from top to bottom, respectively.
Horizontal dotted lines mark the one and three sigma levels.
In the bottom panels solid and dashed lines refer to $S/N>1$ and $S/N>2$, respectively.
The inset in the top panels show the adopted filter, with the dotted vertical line 
marking the position of the Lyman limit at the $z_{min}$. 
}
\end{figure}

We run MC simulations on the area $\sigma_{lens}=10~arcmin^{2}$ by varying 
for each galaxy the parameters involved in Eq.~\ref{eq:f900_obs} 
and derive the ionised flux $F_{\rm LyC}^{Lens}$ as a function of \fesc.
The flux is then compared with the depth of the U-band survey under study.
The procedure is described in details in V10b. Briefly,
a redshift is randomly extracted from the interval {\it dz} (uniformly) and associated
to a galaxy with a given $r$ magnitude extracted randomly from the
magnitude range 26.5--32.5 (accordingly with the magnitude distribution provided 
by the LF). The intergalactic transmission 
$T_{LyC}^{IGM}=e^{-\tau^{\rm IGM}_{\lambda}}$ has been derived from thousands of random realizations 
at the extracted redshift value and convolved with the adopted U-band filter.
The lower limit $z_{min}$ of the redshift range {\it dz} is dictated by the adopted filter shape
that probes LyC at $z>3$, $3.2$ and $3.3$ for the three filters {\it F336W}, {\it U-strom} and {\it LBC-U},
respectively. The ideal upper limit would be $z_{min}+0.1$ such that the 
closest region blueward the Lyman limit is probed. However we relax to $z_{min}+0.5$ by allowing
more sources to be included (in order to reach $N_{spec} \simeq 30$ in the three cases).
The IGM prescription adopted here includes all the intervening absorption systems and
modulates properly the $F_{LyC}^{Lens}$ signal for increasing redshift.
This is also supported by the recent direct detection of LyC at $\lambda_{rest}<830$\AA~(V10b).

The magnification factor ($\mu$) is associated to each galaxy by extracting 
it randomly from the distribution $P(\mu)\sim \mu^{-3}$, as 
explained earlier in the text
(we assume that galaxies are distributed uniformly over the sky).
The \fesc is explored in two different regimes of variability: 
(1) it is fixed to a constant value 
for all $N_{spec}$ sources spanning the range 0.01-1.00, with steps of 0.01 
(i.e., the average $<f_{esc}>=f_{esc}$ of each galaxy) and 
(2) it is fixed to values 0 or 1, assigned randomly such that the 80\% of the $N_{spec}$ 
sample have \fesc=0 and 1 for the remaining sources, i.e., $<f_{esc}>=0.2$ with large
variance (the ``On/Off case'').

In summary, the sample of galaxies derived from the LF in the considered volume 
(area of $10~arcmin^{2}$ and {\it dz}$=0.5$) and magnitude interval $26.5<r<32.5$,
has been extracted 10000 times by varying \fesc, $e^{-\tau^{\rm IGM}_{\lambda}}$, 
$(L_{1500}/L_{LyC})_{\rm int}$, $(F_{1500})_{\rm obs}$ and $\mu$
(no dust has been considered). Figure~\ref{MC} shows the results for the {\it F336W} case.
The resulting median signal to noise ratio 
of the stacked LyC fluxes of the galaxies with $R<26$ 
is shown as a function of \fesc and U-band depth (left panels).
The signal to noise $S/N_{LyC}^{stack}$ is calculated as 
$[\sum_{i=1}^{N_{spec}} F_{\rm LyC}^{i,Lens} / N_{spec}]/[1\sigma_{U} / \sqrt{N_{spec}}]$, 
where $1\sigma_{U}$ is the depth of the U-band survey and $N_{spec}\simeq 30$, i.e., the number of 
surviving LBGs with $R<26$. The number of direct LyC detections above the 2-sigma
limit ($F_{\rm LyC}^{Lens} > 2\sigma_{U}$) is also reported in the same Figure~\ref{MC}
(right panels). While the median $S/N_{LyC}^{stack}$ of the two \fesc distributions,
constant and ``On/Off'', are compatible if $<f_{esc}>=0.2$ 
(as expected from the $S/N_{LyC}^{stack}$ calculation), the variance is larger in 
the ``On/Off'' case. Also the number of direct detections are different between the two, 
because by definition we fix a maximum fraction of detectable sources.
It turns out that with a U-band depth $\gsim 30~(1\sigma)$, the S/N of the stacking 
is $>1(3)$ for \fesc$>0.05(0.18)$. 
The average number of individual detections ($>2\sigma$) is $>3$ for \fesc$>0.10(0.30)$
if the U-band depth is $31(30)$.
These detections are originated by magnification of sources with $r>26.5$ that
migrate from their original magnitude bins to the brighter lensed regime ($R<26$).
The contribution of each bin to the $N_{spec}$ sample is shown as dotted lines in Figure~\ref{MC}.
As a consistency check, and as expected, their number per magnitude bin flattens if the faint 
end slope of the LF is $\alpha=-3$  (i.e., the growth of the number counts and the dimming of the lensing 
cross-section compensate each other). The flattening of the number of direct detections 
as \fesc increases (in the fixed \fesc case, Figure~\ref{MC})
is due to IGM attenuation, i.e., the interception of Lyman limit systems (LLSs) or
damped Lyman-alpha systems (DLAs) along the line of sight suppresses
the ionising flux even for high \fesc values.

Figure~\ref{MC2} shows the same quantities as a function of the U-band depth
by fixing \fesc to 10, 15 and 20\% for all sources. It has been performed for three 
filters: {\it F336W}, {\it U-strom} and {\it LBC-U}.
Clearly, the deeper U-band surveys give the best constraints to
the faint galaxies, in particular those with mag-U$>30$ are the more effective
and increase rapidly as the mag-U$=31$ limit is approached. The intermediate band 
filter like the u-str\"omgren is an optimal solution.
It is worth noting that observing a massive cluster with $\theta_{E} \simeq 1~arcmin$
(such us MACS 0717.5+3745, see below), we expect $\sim$ 10 galaxies with $R<26$
and $r>26.5$. If all of them have \fesc = 15\%, then with a U-band (u-str\"omgren-like) 
depth of 30.5 we expect to detect 3 of them at $S/N>1$ and a $S/N_{LyC}\simeq 3.0$ for the stacking.
Therefore, we expect to observe signal already with a single deep pointing (of $\sim$60hr) 
with a 8-10$m$ class telescope.

\section{The current observations}

As mentioned above, the interception of LLSs or DLAs
makes IGM attenuation quite stochastic and severe for ionising radiation
at $z\gsim3$, therefore the observations of magnified galaxies along different lines
of sight are necessary to attenuate its effect.
Assuming these absorbers are associated with individual structures with physical
size of $\lsim 50$ kpc (physical) (e.g., Yajima et al. 2012), 
different lines of sight separated by more 
than $\sim 10~arcsec$ will have a relatively small transverse correlation among
high column density absorbers, even along a single cluster.

Currently, there are in the literature several massive
clusters of galaxies characterised by relatively
large lensing cross-sections, e.g., $\theta_{E} \sim 0.7-0.9~arcmin$, 
like MACS 0717.5+3745 (Zitrin et al. 2009),
1E0657--56 (''Bullet Cluster'', Bradac et al. 2009) and A1689 (Broadhurst et al. 2005).
Zitrin et al. (2012) recently measured the Einstein radius of 10000 galaxy clusters 
from the SDSS.
In particular they identified $\simeq$40 candidates with $\theta_{E} > 0.7~arcmin$. The largest Einstein 
radius was $\theta_{E} \simeq 1.1~arcmin$ for the most massive cluster (for a source at $z_{s}=2$).
The MACS survey (MAssive Cluster Survey) also provides a 
statistically complete sample of very X-ray luminous distant clusters of galaxies 
($0.3 < z < 0.7$, Ebeling et al. 2010).
Therefore, given the existing observations, the area of $\sigma_{lens}=10~arcmin^{2}$ 
considered in this letter can be easily achieved with 5 clusters.
It is worth mentioning the ongoing Cluster Lensing And Supernova
survey with Hubble (CLASH), a multi-treasury program that, in addition
to other available ground-based observations is observing (524 HST
orbits) 25 X-ray selected massive galaxy clusters with new HST
panchromatic imaging capabilities (16 filters): the ACS and both the
UVIS and IR channels of the WFC3.  While 20 clusters were chosen to be
X-ray selected, relatively relaxed clusters with no lensing-selection
bias, the 5 additional were chosen to be high-magnification clusters
with $\theta_{E}>0.6~arcmin$ (Postman et al. 2012).  
Even though the survey was not designed to provide deep limits on LyC
for faint galaxies, 
the homogeneous panchromatic photometry on a
statistically significant sample of massive clusters, together with
magnification maps and high S/N spectroscopy on a number of highly
magnified galaxies at $z\gsim 3$, will yield a very valuable testbed
for probing LyC emission at the faint end of the galaxy LF.

\section{Conclusions}
We have explored the possibilities offered by gravitational lensing
in constraining the ionizing emission from faint galaxies at $z > 3$,
which has a great impact on studies of the reionization
even at significantly higher redshifts ($z > 6$).
We have shown that with current ground-based
facilities and the magnification provided by known
clusters of galaxies it is possible to constrain the \fesc quantity
down to 10-15\% for faint luminosities ($L\lsim 0.1L^{*}$).
In particular, strong lensing provides the opportunity to make a significant
step forward in this field in the pre-ELT era:
{\bf (1)} The magnification allows us to measure redshifts for
very faint galaxies, i.e., to fix the wavelength position of
the Lyman limit and put strong constraints on the LyC with
U-band imaging for galaxies down to $L\sim  0.05 −- 0.1L^{*}$.
{\bf (2)} The magnification allows us to perform accurate
spatial analysis of the LyC emission (if detected), down to
hundreds of parsecs.
If, on the one hand, the magnified area
increases the probability of intercepting a foreground lower-z
source that might mimic LyC emission (Vanzella et al. 2010a),
on the other hand, the possible presence of multiple images
of the same background source will help to solve the problem,
since a real LyC detection will be present in all the counterparts
(if $\mu$ are similar).
The transverse separation of the light paths decreases rapidly as the source
redshift is approached. Therefore the IGM attenuation is
practically the same for all the multiple images.
{\bf (3)} It has been suggested that the \fesc increases with
decreasing luminosity (e.g., Yajima et al. 2011) and the ionizing
background is mainly produced by a large number of sub$-L^{*}$ 
galaxies. A nice feature of the present approach is to put
a limit to the process of attributing the ionizing contribution
to lower and lower luminosities with the increasing observational
depth of null results. This in fact would imply a
steepening of the faint end of the LF that in principle can
be excluded by the increasing number of highly magnified
sources expected to be detected through lensing.

We have shown that deep U-band observations 
(as in V10b) of five massive galaxy clusters provide 
a sample of intrinsically faint ($L\lsim0.1L^{*}$) and
magnified $z~\sim 3$ galaxies (about 30) useful to
investigate \fesc down to 10-20\%.
An increasing number of well studied clusters for this study, with 
multi-band HST photometry, spectroscopy and lens modeling, are 
becoming available as part of the CLASH project.

\section*{Acknowledgements} 
We thank the anonymous referee for the fruitful comments.
We thank G. Cupani and F. Fontanot for useful discussions.
AZ is supported by the ``International Spitzenforschung II/2'' of the
Baden-W\"urttemberg Stiftung.
We acknowledge financial contribution from the agreement ASI-INAF I/009/10/0
and from the PRIN MIUR 2009 ``The Intergalactic Medium as a probe of the 
growth of cosmic structures'' and the Dark Cosmology Centre which is funded 
by the Danish National Research Foundation. 


\end{document}